\documentstyle[preprint,revtex]{aps}
\begin{document}
\draft
\pagestyle{empty}
\null\vskip -.8in
\centerline{June 1992 \hfill} \vskip -4mm
\preprint{NUHEP-TH-92-13}
\preprint{UM-P-92/49, OZ-92/16}
\preprint{Zu-Th 16/92}
\preprint{DOE/ER/4051-061-INT92-07-04}
\begin{title}
Neutron electric dipole moment due to Higgs exchange\\
in Left-Right symmetric models
\end{title}
\author{D. Chang$^{1}$, Xiao-Gang He$^{2}$, W.-Y. Keung$^{3}$,\\
B.H.J McKellar
$^{4}$, and D. Wyler$^{5}$}
\begin{instit}
$^1$ Department of Physics and Astronomy\\
Northwestern University, Evanston, IL 60208, USA\\
$^2$ Research Center for High Energy Physics, School of Physics\\
University of Melbourne, Parkville, Vic. 3052 Australia\\
$^3$ Physics Department, University of Illinois at Chicago, IL 60680, USA\\
$^4$ Institute for Nuclear Theory, University of Washington,
Seattle, WA 98105 USA\\
$^5$ Theoretische Physik, Universit$\ddot{a}$t Z$\ddot{u}$rich, 8001
Z$\ddot{u}$rich,
Switzerland
\end{instit}
\vskip -.3 in
\begin{abstract}
In this paper we study  the
neutron electric dipole moment (EDM) due to Higgs
boson exchange in Left-Right symmetric models.
In pseudo-manifest Left-Right symmetric
models, the neutral Higgs contribution is
smaller than that from the charged Higgs.
The charged Higgs contribution at the two loop level can be as
large as the  experimental upper bound.
In non (pseudo) manifest Left-Right symmteric
models,
the neutral Higgs exchange contribution can reach
the experimental upper bound. The Higgs exchange contributions can be more
important than
the ones from W-boson
exchange  due to $W_L - W_R$ mixing.
\end{abstract}
\vskip -.3 in
\vskip -.3 in
\newpage
\pagestyle{plain}
\narrowtext
One of the outstanding problems of particle physics today is the origin of
CP violation. CP violation has only been observed in the neutral kaon
system, and many models have been proposed to explain it \cite{cp}.
In order to determine the source (or sources) responsible for CP violation,
it is important to find other processes which also violate CP.
The measurement of the neutron EDM, $D_n$, is a very promising area of
investigation.
A very stringent
upper bound on the neutron EDM ($D_n$) has been obtained \cite{1}, $|D_n|\;
<\;1.2\times 10^{-25}\;$ \hbox{ecm}, whereas
the standard model \cite{2} predicts a very small $D_n$
($<\; 10^{-31}\;\hbox{ecm}$).
There are similarly stringent bounds on the electron \cite{e}
 and atomic \cite{a} EDMs.
Assuming that the strong CP $\theta$ parameter is negligible, if a
neutron EDM within five orders of magnitude of
the experimental upper bound should
be detected, it signals physics
beyond the standard model. In extensions of the
standard model it is indeed
possible to have a large neutron EDM \cite{3,4}.
CP violation due to Higgs exchange is an example
of such models. Recently, several authors have exploited some new classes
of two
loop diagrams which induce a large neutron EDM \cite{5,6,7,8,h}. In this paper
we study
these new contributions due to Higgs exchange in Left-Right symmetric models
and compare them with the contributions from W-boson exchange due to
$W_L -W_R$ mixing
\cite{8,9,bclp,10}.
The neutron EDM due to Higgs exchange at the one loop level in Left-Right
symmetric models has
been considered before \cite{11}.
Here we will discuss both the one loop
and two loop contributions.

The gauge group of the Left-Right symmetric models is
$SU(3)_C\times SU(2)_L\times SU(2)_R\times U(1)_{B-L}$ \cite{lr}.
Under this group the left and right handed fermions transform as
\begin{eqnarray}
Q_L = (3,\;2,\;1,\; 1/3)\;,\;\;&Q_R = (3,\;1,\;2,\;1/3)\;,\nonumber\\
L_L = (1,\;2,\;1,\; -1)\;,\;\;&L_R = (1,\;1,\;2,\;-1)\;,
\end{eqnarray}
where $Q$ and $L$ are quarks and leptons respectively.
In order to give fermion masses through the tree level Higgs-fermion couplings,
at least one bi-doublet representation of Higgs boson,
transforming as $\phi = (1,\;2,\;2,\;0)$, is needed. It can be written as
\begin{equation}
\phi =\left(\begin{array}{ll}
\phi^0_1\;\;&\phi_1^+\\
\phi_2^-\;\;&\phi_2^0
\end{array}\right )\;,
\end{equation}
and its vacuum expectation value (VEV) is
\begin{equation}
\langle \phi\rangle  =\left(\begin{array}{ll}
v_1\;\;\;&0\\
0\;\;\;\;&v_2e^{i\delta}
\end{array}\right )\;.
\end{equation}
In this notation, $\phi$ transforms as $U_L \phi U_R^{\dagger}$ under
$SU(2)_L \times SU(2)_R$.
In order to break $SU(2)_R$ at a higher scale,
additional Higgs representations are needed. There are two
traditional ways of introducing these Higgs representations,
\begin{eqnarray}
a)&\;\;H_L = (1,\;2,\;1,\;1)\;,\;\;&H_R = (1,\;1,\;2,\;1)\;;\nonumber\\
b)&\;\;\Delta_L = (1,\;3,\;1,2),\;\;&\Delta_R = (1,\;1,\;3,\;2)\;.
\end{eqnarray}
In case a) neutrinos can have only Dirac masses.  In case b) neutrinos can
have both Dirac and Majorana masses and the lighter neutrinos have naturally
small masses due to the see-saw mechanism.  For this reason, the case b) is
usually favored in the literature.  However, for our purposes, the two cases
result in similar phenomenology.
If the VEV of $\langle H_R\rangle\;(\langle \Delta_R\rangle)  = v_R$ is
larger than $v_1$, $v_2$ and
the VEV of $\langle H_L\rangle\; (\langle\Delta_L\rangle)  = v_L$
, the symmetry breaking
scales for $SU(2)_L$ and $SU(2)_R$ are well separated.
If $v_1v_2 \neq 0$, there is a
mixing between $W_L$ and $W_R$ with a mixing angle
$\zeta\approx
v_1v_2/v_R^2$ for a) and $2v_1v_2/v_R^2$ for b).
In the following, we shall adopt case a) for illustrative purpose whenever we
need to.
For simplicity we will assume $v_L = 0$.
In order to make this assumption consistently, it is necessary to
impose additional discrete symmetries to eliminate the terms linear in
$H_L$ in the Higgs potential \cite{note}.

The Higgs-quark couplings are given by
\begin{eqnarray}
L_Y &=& \bar Q_L f\phi Q_R + \bar Q_L h \tau_2 \phi^*\tau_2 Q_R + H.C.\;,
\end{eqnarray}
where $f$ and $h$ are $3\times3$ matrices.
We obtain the mass matrices for quarks
\begin{equation}
M'_u = fv_1 + hv_2 e^{-i\delta}\;, M'_d = fv_2e^{i\delta}+hv_1\;,
\end{equation}
which can be diagonalized by the following transformation:
\begin{equation}
M'_u = V_L^{u\dagger}M_u V^u_R\;,\;\; M'_d = V_L^{d\dagger}M_dV^d_R\;,
\end{equation}
where $M_{u,d}$ are the diagonalized mass matrices for up and down quarks
respectively.
The mixing matrices for the charged currents are
\begin{equation}
V_L = V^u_LV^{d\dagger}_L\;,\;\;V_R = V^u_RV^{d\dagger}_R\;.
\end{equation}

In general $V_L$ and $V_R$ are independent. One can always parametrize $V_L$
in the conventional way in which there is only one CP violating phase for three
generations of quarks. Then, in general, $V_R$ will have six CP violating
phases. In special cases, the number of CP violating phases is reduced.
For simplicity, we shall impose the
following Left-Right exchange symmetry, S:
\begin{equation}
Q_L\; \leftrightarrow\; Q_R\;,\;\; \phi\;\leftrightarrow\; \phi^\dagger\;,
\end{equation}
on the Lagrangian.
It implies $f=f^\dagger$ and $h = h^\dagger$.
In the following we shall consider three cases.
\begin{enumerate}
\item
CP is broken explicitly, however $\delta=0$.
In this case the mass matrices
are hermitian and can be diagonalized by unitary tranformations.
Therefore we have
\begin{eqnarray}
V_L = V_R\;.
\end{eqnarray}
We shall refer to this case as the manifest Left-Right (MLR) symmetric case.
Since the phases in $V_L$ and $V_R$ can be simultaneously removed, we can
assume that both are transformed into Kobayashi-Maskawa (KM) form.
\item    CP is assumed to be spontaneously broken.
In this case, $f$ and $h$ are real and symmetric but $\delta\neq 0$.
To diagonalize a symmetric matrix it is possible to use $V_L^* = V_R$ in the
bi-unitary transformation.  Therefore in arbitrary basis
one would have \cite{mps}
\begin{eqnarray}
V_R = J_uV_L^*J^*_d\;. \label{vlr}
\end{eqnarray}
with
\begin{eqnarray}
J_u = \hbox{diag} (e^{-i\alpha_u}\;,\;e^{-i\alpha_c}\;,\;e^{-i\alpha_t})\;,\;\;
J_d = \hbox{diag} (e^{-i\alpha_d}\;,\;e^{-i\alpha_s}\;,\;e^{-i\alpha_b})\;.
\end{eqnarray}
We shall refer to this case as the pseudo-manifest Left-Right (PMLR) symmetric
case.  We shall take the basis in which $V_L$ is in KM form.
\item  CP is explicitly broken and $\delta$ is also nonzero.  In this case
there is no simply relation between $V_L$ and $V_R$.  If one also does not
insist on
the S symmetry of Eq.(9), $V_L$ and $V_R$ are completely independent. We
refer to this case as
the non-manifest Left-Right (NMLR) symmetric case.
An interesting special case of this which produces interesting
phenomenological consequences is one in which $V_R$ can be written as \cite{oe}
\begin{eqnarray}
V_R = \left(\begin{array}{lll}
1\;\;\;&0\;\;\;&0\\
0\;\;\;&V_{Rcs}\;\;\;&V_{Rcb}\\
0\;\;\;&V_{Rts}\;\;\;&V_{Rtb}\end{array}\right)
\end{eqnarray}
This form maximizes the effect of the flavor changing neutral Higgs as we shall
show later.
\end{enumerate}

In order to study Higgs contributions to the neutron EDM,
we need to find out the physical Higgs couplings to quarks.
For simplicity we will choose case a) of Eq.(4)
and assume that CP is broken spontaneously in case 2) or explicitly in case 3)
from now on.
In that case, there is one charged Higgs eigenstate $\chi^+$ which couples
directly to quarks \cite{12},
\begin{equation}
\chi^+ = {1\over T}\left[(v_1^2-v_2^2)H_R^+ + v_R(v_1\phi_1^+ +v_2e^{i\delta}
\phi^+_2)\right]\;.
\end{equation}
where $T^2 = v^2v_R^2 + (v_1^2 - v_2^2)^2$, $v^2 = v_1^2 +v_2^2$.
The charged Higgs boson associated with $H_L$ does not
mix with the others because
of the discrete symmetry \cite{note} and does not couple to fermions at all.
There are three physical neutral Higgs bosons which couple to quarks.
We analyze in a convenient basis, $\phi'^0_1$ and $\phi'^0_2$,
which are linear combinations of $\phi^{0*}_1$ and $\phi^0_2$ such that
$\langle \phi'^0_1 \rangle \neq 0$ and $\langle \phi'^0_2 \rangle =0$.
The physical neutral Higgs bosons are then
expressed as linear combinations of $H_1$, $H_2$ and $H_3$.
Here $H_1$ is the the real part of $\phi'^0_1$ while $H_2$, $H_3$ are real and
imaginary parts of $\phi'^0_2$.  They can be written explicitly as \cite{12}
\begin{eqnarray}
H_1&=&  \cos\theta \ \phi_{1R} + \sin\theta \cos\delta \ \phi_{2R}
       + \sin\theta \sin\delta \ \phi_{2I}\;,\nonumber\\
H_2&=& -\sin\theta \ \phi_{1R} + \cos\theta \cos\delta \ \phi_{2R}
       + \cos\theta \sin\delta \ \phi_{2I}\;,\\
H_3&=&  \sin\theta \ \phi_{1I} - \cos\theta \sin\delta \ \phi_{2R}
       + \cos\theta\cos \delta \ \phi_{2I}\;\nonumber.
\end{eqnarray}
where $\cos\theta =v_1/v$ and $\sin\theta = v_2/v$, and $\phi_{iR,L}$ denote
the real and imaginary parts of $\phi^0_i$ respectively.

For case b) of Eq.(4), the situation is more complicated because it is
harder to eliminate the term linear in $\Delta_L$ \cite{cm}.
If these terms remain
then $\langle \Delta_L\rangle \neq 0$ and
the singly charged Higgs boson in $\Delta_{L}$ will also mix
with $\phi^+_i$ just as $\Delta_R$ does.
The neutral components of $\Delta_{L,R}$ will also mix with $H_i$
defined in Eq.(15) \cite{k}. However,
if $v_R \gg v_i$ these mixings will be small.
The dominant components which couple to quarks are still
$\chi^+ \approx \cos\theta
\phi^+_1 + \sin\theta e^{i\delta}\phi^+_2$ and $H_i$ just as in the case a).

The neutral Higgs bosons defined in Eq.(15) are in general not mass
eigenstates.  However in order to simplify the discussion, we will take these
particles to be mass eigenstates in the following
for PMLR and NMLR models.  In these two cases, the mixings in Eq.(15)
already reflect the full complexity of the problem as far as the CP violating
phenomenology is concern.
If CP is explicitly broken in the Higgs self couplings,
as is required in the case of MLR models (since $\delta =0$),
the mixings between these neutral Higgs bosons
are more complicated and important.  We will comment on this later.
The Yukawa interactions of these Higgs bosons to the quark sector are
\begin{eqnarray}
L_{Yukawa} = {(\sqrt 2 G_F)^{1/2}\over \cos2\theta}&\Bigl\{ &
     \sqrt{2}\Bigl[\bar U_L(M_uV_R - V_LM_d \sin2\theta e^{-i\delta})D_R
                                                                \nonumber\\
           &-& \bar U_R(V_RM_d - M_uV_L\sin2\theta
                       e^{-i\delta})D_L\Bigr]\chi^+        \nonumber\\
&+& \bar U_L M_u (\cos2\theta H_1 -\sin2\theta H_2 + i\sin2\theta H_3)U_R\\
&+& \bar D_L M_d(\cos2\theta H_1 - \sin2\theta H_2 - i\sin2\theta H_3)D_R
                                                     \nonumber\\
&+&\bar U_L(V_LM_dV^\dagger_R)e^{-i\delta}(H_2 - iH_3)U_R \nonumber\\
&+&\bar D_L (V_L^\dagger M_u V_R)e^{i\delta} (H_2 +iH_3)D_R \Bigr\}+ H.c.\;
                                                      \nonumber
\end{eqnarray}
One should note that in contrary to the multi-doublet extensions of Standard
Model frequently discussed in the literature \cite{6} the charged Higgs boson
$\chi^+$ has right-handed couplings $M_u V_R$ that are proportional
to up-type quark masses, in addition to
the usual left-handed ones.
In particular, these new couplings depend on the $V_R$
mixing matrix which is not severely constrained experimentally.  Therefore the
$d_R$ quark can in principle have a large mixing with $t_L$ through charged
Higgs boson.  This fact has been observed before \cite{11} but has not been
emphasized.  Similarly, in the last
term the the neutral Higgs couplings is also
proportional $M_u V_R$.  They are partly responsible for the large CP violating
effects that we shall discuss later.

We will use the standard KM convention \cite{book} for $V_L$ with
Im $V_{Ltb}=0$.
We also set $|V_{us}|_{L,R}\approx |V_{us}|_{L,R} = 0.22$, $|V_{td}|_{L,R}
= 0.006$, in PMLR models. In NMLR models,
$V_{Rij}$ can be different from $V_{Lij}$.  We shall assume it is of the form
in Eq.(13) to maximize the effect of CP violation.
For the quark masses we will use:
$m_u(1\;\hbox{GeV}) = 4.2 \hbox{MeV}$, $m_d(1\;\hbox{GeV}) = 7.5 \hbox{MeV}$,
$m_s(1\;\hbox{GeV}) = 150  \hbox{MeV}$, $m_c(m_c) = 1.4 \hbox{GeV}$,
$m_b(m_b) = 5 \hbox{GeV}$ and $m_t(m_t) = 150 \hbox{GeV}$.
Since $m_t \gg m_b$, a natural value for $\theta$ is
$\sin2\theta \approx 2{m_b\over m_t}$.

The $H_1$ boson behaves like the Higgs boson of the Standard Model.
Its coupling does not mediate flavour changing
neutral currents (FCNC) and does not violate CP at the tree level.
But $H_2$ and $H_3$ do both.
Note that in the usual multi-doublet extensions of the Standard Model, such
FCNC-mediating Higgs bosons can be avoided by introducing
a discrete symmetry \cite{wg}.
However they
are  essential parts of the usual Left-Right Symmetric Models \cite{bclp}.
Therefore, in this case, instead of trying to avoid them, we shall investigate
under what circumstances their effect can be large and detectable.
Because $H_2$ induces FCNC at the tree level,
its mass must be sufficiently large in order not  to yield a too large
mass difference between $K_L$ and $K_S$.
This consideration constrains the mass of $H_2$ to be larger than
$8\;\hbox{TeV}$ \cite {14} in the
MLR and PMLR models.
In PMLR models with spontaneous CP violation, it was difficult to get
$\delta \neq 0$ if one used only minimal Higgs multiplets \cite{lw}.
However it was also observed \cite{lw} that such solution can indeed be
obtained if one is willing to make a slight extension of Higgs sector.
The lower bound on the mass derived from the absence of FCNC only
applies to neutral Higgs bosons.  In our
estimates, for PMLR models we will use $10\;\hbox{TeV}$
for neutral Higgs mass. The charged Higgs $\chi^+$ can have a smaller mass.
When $V_L$ and $V_R$ are independent from each other,
if one takes the special form of $V_R$ in Eq.(13), the experimental
lower bound for the $H_2$ mass can be smaller.

We are now ready to estimate the Higgs contributions to the neutron EDM.
We shall consider the following three interactions which can give
important contributions to the neutron EDM,
\begin{eqnarray}
\hbox{the quark edm},\qquad  O^\gamma &=&{} -{d_q\over 2}i\bar q
\sigma_{\mu\nu}\gamma_5 F^{\mu\nu} q\;,\nonumber\\
 \hbox{the quark color
edm},\qquad O^C_q & =&{} -{f_q\over 2}ig_s\bar q
\sigma_{\mu\nu}\gamma_5G^{\mu\nu}q\;,\\
 \hbox{the gluon color edm},\qquad O^C_g &=&
{}-{1\over 6} Cf_{abc}G^a_{\mu\nu}G^b_{\mu\alpha}\widetilde G^c_{\nu\alpha}\;,
\nonumber
\end{eqnarray}
where $F^{\mu\nu}$ is the photon field strength,
$G^{\mu\nu}$ is the gluon field strength and $\tilde G^{\mu\nu} = {1\over
2}\epsilon_{\mu\nu\alpha\beta}G^ {\alpha\beta}$.

There are many ways to estimate the contributions of these operators to the
neutron electric dipole moment, $D_n$.
Using $SU(6)$ relations we
have \cite{3} \begin{equation}
D_n(d_q) = {1\over 3}(4d_d - d_u)\;,\;\;
D_n(f_q) = {1\over 3}\left({4\over 3}f_d+ {2\over 3}f_u\right)e\;.
\end{equation}
The estimate for $O^C_q$ is more uncertain than that of
$O^\gamma$.  Various other estimates\cite{7}
and calculations
using sum rule techniques\cite{kw} give a range between 0.05 and 1
for the ratio $D_n(f_q)/ef_q$. A recent reevaluation\cite{ref:kk}
confirms in fact the result of Eq.(18).
For the contribution from $O_g^C$, we
use the naive dimensional analysis (NDA) to
estimate the neutron EDM \cite{5}
\begin{equation}
D_n \approx  {eM\over 4\pi} C\;,
\end{equation}
where $M=4\pi f_\pi= 1190 \hbox{MeV}$ is the scale of chiral symmetry breaking.
An alternative estimate using QCD sum-rules \cite{bu} gives a value smaller by
about a factor of 30.  The sum-rule result involves additional assumptions such
as $\eta$ dominance and its reliability is hard to assess.  However, the NDA
estimate is also plagued by uncertainties,
in this case an arbitrary assumption about the normailzation.
The comparison of these two estimates may be used
as an estimate of the uncertainty in the
calculation of hadronic matrix elements.

A non zero--value for $f_s$ will also generate a neutron EDM.
It was estimated to give \cite{h}
\begin{equation}
D_n(f_s) \approx 0.03f_s e\;.
\end{equation}
As we will show later, in some scenarios, $f_s$ can give rise to the dominant
contribution.

In models of CP violation, the quark edm, $d_q$,
and the quark color edm, $f_q$, can be generated at the one
and two loop levels.
The gluon edm, $O^C_g$, are typically generated at the two loop level.
The one loop contribution to $d_d$ and $f_d$ from the neutral Higgs boson,
as shown in Fig.~1, is given by \cite{11}
\begin{eqnarray}
d_d &\approx& \left(-{1\over 3}e\right){m_bG_F\over 8\sqrt 2
\pi^2}{m_t^2\over \cos^2   2\theta m_H^2}
\ln\left({m_H^2\over m_b^2}\right)\eta_d \hbox{Im}
\left(V_{Ltd}^*V_{Rtb}V_{Ltb}^*V_{Rtd}
e^{2i\delta}\right)\;,\\
e f_d &\approx&- 3{\eta_f\over \eta_d} d_d\;,\nonumber
\end{eqnarray}
Note that it is assumed that the neutral Higgs couplings is dominated by the
flavor changing neutral current, the last term in Eq.(16).
In PMLR models, using Eqs.(11,12),
\begin{equation}
\hbox{Im}\left(V_{Ltd}^*V_{Rtd}V_{Rtb}V_{Ltb}^*e^{i2\delta}\right) \approx
|V_{td}|^2\sin(\alpha_d + \alpha_b  + 2\delta - 2\alpha_t )\;.
\end{equation}

For the charged Higgs contribution in Fig.~2, we obtain \cite{11}
\begin{eqnarray}
d_d &\approx& ({2\over 3}e){m_tG_F\over 4\sqrt 2 \pi^2}
\sin2\theta {m_t^2\over \cos^2
2\theta m_\chi^2} \ln\left({m_\chi^2\over m_t ^2}\right)\eta_d
\hbox{Im}(V_{Ltd}V_{Rtd}^* e^{-i\delta})  \;,\\
e f_d &\approx&{3\over 2}{\eta_f\over \eta_d}d_d\;.\nonumber
\end{eqnarray}
In PMLR models,
\begin{equation}
\mbox{Im}(V_{Ltd}V_{Rtd}^*) = |V_{td}|^2\sin(\alpha_t - \alpha_d -\delta)\;.
\end{equation}
In Eqs.(21,23), $\eta_d = \left({\alpha_s(m_t)\over
\alpha_s(\mu)}\right) ^{16/23}$ and $\eta_f =
\left({\alpha_s(m_t)\over \alpha_s(\mu)}\right)^{14/23}$ are the QCD correction
factors \cite{7}.
Note that $d_d$ is more suppressed by the QCD correction than $f_d$.
Following Ref.\cite{5}, we will use $\alpha_s(\mu) = {4\pi\over 6}$ and
$\alpha_s(m_t)=0.1$.
As we commented before, Eq.(23) is characterized by its
$m_t^3$ dependence, a feature
which distinguishes it from the the usual multi-doublet models.

Using the numerical values quoted before for the parameters,
we find the contribution to $D_n$ from neutral Higgs exchange
to be less than $10^{-28}\hbox{ecm}$ with $m_H = 10 $ TeV.
Using the same parameters
for the charged Higgs boson contribution in PMLR models, we have
\begin{eqnarray}
D_n(d_d) = \left \{ \begin{array}{ll}
3\times10^{-28} \sin(\alpha_t-\alpha_d -\delta)\, \hbox{ecm}\;,\;\;&
m_\chi = 10 \hbox{TeV},\\
1.3\times 10^{-26} \sin(\alpha_t-\alpha_d -\delta)\, \hbox{ecm}\;,
\;\;&m_\chi = 1
\hbox{TeV}\;, \end{array}\right.
\end{eqnarray}
where we have set $\sin2\theta\approx 2{m_b\over m_t} \approx
0.04$. We see that the one loop level Higgs contributions to the
neutron EDM are small. Of course if the mass of the charged Higgs
is much lower than $1\;\hbox{TeV}$,
it is possible to have a larger neutron EDM.
A similar contribution also comes from $f_d$ (about 60\% of $d_d$
contribution).
The contributions from $d_u$ and $f_u$ are smaller
because the couplings are smaller.

The contribution to the neutron EDM from $f_s$ due
to the neutral Higgs boson is given by
\begin{eqnarray}
D_n(f_s) &\approx& 0.03 f_s e\nonumber\\
&\approx& 0.03e {m_bG_F\over8\sqrt 2 \pi^2} {m_t^2\over \cos^22\theta m_H^2}
\ln
\left({m_H^2\over m_b^2}\right) \eta_f \hbox{Im}
\left(V_{Lts}^*V_{Rts}V_{Rtb}V_{Ltb}^*e^{2i\delta}
\right)\\
&=& 2\times 10^{-26}\hbox{ecm} \times  {1\over (0.04)^2}
\hbox{Im}\left(V^*_{Lts}V_{Rts}V_{Rtb}V^*_{Ltb}
e^{2i\delta}\right)\;,\;\;\; m_H = 1 \hbox{TeV}\;.\nonumber
\end{eqnarray}

There is also a similar contribution from the charged Higgs boson. We have
\begin{eqnarray}
D_n(f_s) &\approx& 0.03e {m_tG_F\over 4\sqrt 2 \pi^2}\sin2\theta
{m_t^2\over \cos^2 2\theta
m^2_\chi } \ln\left({m^2_\chi\over m^2_t}\right) \eta_f
\hbox{Im}\left(V_{Lts}V_{Rts}^*e^{-i\delta}
\right)\nonumber\\
&=& 2.7 \times 10^{-26} \hbox{ecm}{1\over (0.04)^2}
\hbox{Im}\left(V_{Lts}V^*_{Rts}e^{-i\delta}
\right)\;,\;\;\; m_\chi = 1\;\hbox{TeV}\;.
\end{eqnarray}

In the special case of Eq.(13), $|V_{Rts}|$ can be larger than
$|V_{Lts}| \sim 0.04$, and therefore these contributions can be near the
experimental upper bound. In PMLR models,
$|V_{Rts}| = |V_{Lts}|$ and the neutral Higgs masses are around $10
\hbox{TeV}$.
 Then,
only Eq.(27) contributes significantly, with values near those in Eq.(25).

We now turn to the two loop contributions.  Once again in this case one can
take advantage of the fact that CP violating neutral Higgs couplings can all be
proportional to $m_t$ instead of having at least one of them proportional to
$m_b$ as in the case of the multi-doublet extensions of Standard Model.
At this level, the neutral Higgs
exchange in Fig.~3 will generate a quark color edm $f_q$ which is
given by \cite{7}
\begin{eqnarray}
f_q &=& {G_F\over 16\sqrt 2 \pi^3}m_q \alpha_s(\mu)\left({\alpha_s(m_t)\over
\alpha_s(\mu)}\right)^{37/23} G({m_t^2\over m_H^2},q)\;,\nonumber\\
G(z,u) &=& f(z)\hbox{Im} Z_{tu} + g(z)\hbox{Im}Z_{ut}\;,\\
G(z,d) &=& f(z)\hbox{Im} Z_{td}+g(z)\hbox{Im}Z_{dt}\;.\nonumber
\end{eqnarray}
For $z\; \ll 1$,
\begin{equation}
f(z) \approx g(z) \approx {1\over 2}z(\ln z)^2\;,
\end{equation}
where $\hbox{Im} Z_{ij}$ are defined through
\begin{equation}
\hbox{Im} Z_{ij} = 2\gamma_i\beta_j\;.
\end{equation}
with
\begin{eqnarray}
L_{int} &=& (2\sqrt 2 G_F)^{1/2} (m_t\gamma_t\bar t t + im_t \beta_t \bar t
\gamma_5 t + m_d \gamma_d\bar d d \nonumber\\
&+& i m_d \beta_d \bar d\gamma_5 d + m_u \gamma_u
\bar uu + i m_u\beta_u \bar u \gamma_5 u)H_2\;,
\end{eqnarray}

In PMLR models the largest contribution to
$f_d$ is from the term
proportional to Im$Z_{td}$, we have
\begin{equation}
\hbox{Im}Z_{td} =- {\sin2\theta\over \cos^22\theta} \,\hbox{Im}
\left((V_{Lud}^*V_{Rud}
+{m_c\over m_s}V_{Lcd}^*V_{Rcd} + {m_t\over m_d}V_{Ltd}^*V_{Rtd})e^{i\delta}
\right)\;;
\end{equation}
and
\begin{equation}
\hbox{Im} Z_{td} \approx -{m_c\over m_d} {|V_{cd}|^2\over
      \cos^2 2\theta}\sin 2\theta
\sin(\alpha_d-\alpha_c +\delta)\;.
\end{equation}
The contribution to $D_n$ is again small,
$D_n < 4\times 10^{-29} \;\hbox{ecm}$ for $m_H = 1 \hbox{TeV}$.
The $f_u$ contribution is even smaller.

In the special case of Eq.(13), the contribution from $f_s$ again
dominates over other contributions. Changing the subscript d to s in equations
 (28) and (32), we obtain $f_s$.
The resulting value of the neutron EDM is given by
\begin{eqnarray}
D_n(f_s) &\approx& 0.03f_s e\nonumber\\
&\approx& 2\times 10^{-27} \hbox{ecm}\,
\hbox{Im}\left(V_{Lcs}^*V_{Rcs}e^{i\delta}+
{m_t\over m_c}V_{Lts}^*V_{Rts}e^{i\delta}\right)\;, m_H = 1 \hbox{TeV}\;.
\end{eqnarray}
This contribution is small.

The operator $O^C_g$ will also be generated at the two loop level. We find the
neutral Higgs contribution to $D_n$ through this mechanism
to be \cite{5}
\begin{eqnarray}
D_n &\approx& e\xi M {\sqrt 2 G_F\over (4\pi)^2}\,\hbox{Im} Z_{tt} h({m_t^2
\over m_H^2})
\;,\nonumber\\
\xi &=& \left({g(\mu)\over 4\pi}\right)^3\left({\alpha_s(m_b)\over
\alpha_s(m_t)}\right)^{-54/23} \left({\alpha_s(m_c)\over
\alpha_s(m_b)}\right)^{-54/25}\left({\alpha_s(\mu)\over \alpha_s(m_c)}
\right)^{-54/27} \approx 6\times 10^{-5}\;. \end{eqnarray}
For $z \ll 1$,
\begin{equation}
h(z) \approx {1\over 2}z\ln z\;.
\end{equation}
We have
\begin{equation}
\hbox{Im} Z_{tt} = -{m_b\over m_t}{\sin2\theta\over \cos^22\theta}
\,\hbox{Im}(V_{Ltb}V_{Rtb}^*
e^{-i\delta})\;.
\end{equation}
This effect is extremely small $D_n < 10^{-30} \hbox{ecm}$.
In the special case of Eq.(13), this contribution can be larger ($\sim
10^{-28} \hbox{ecm}$)
because the neutral Higgs mass is less constrained.

The charged Higgs contribution in Fig.~4 to the neutron EDM
via the operator $O^C_g$ is
give by \cite{5}
\begin{eqnarray}
D_n &\approx& e\xi' M {\sqrt 2 G_F
\over (4\pi)^2}\, \hbox{Im} Z' h'\left({m_t^2\over
m_H^2}\right)\;,\nonumber\\ \xi' &=& \left({g_s(\mu)\over 4\pi}\right)^3
\left({\alpha_s(m_b)\over \alpha_s(m_t)}\right)^{-14/23}
\left({\alpha_s(m_c)\over \alpha_s(m_b)}\right)^{-54/25}\left({\alpha_s(\mu)
\over
\alpha_s(m_c) }\right)^{-54/27} \approx 3 \times 10^{-4}.
\end{eqnarray}
For $z \ll 1$,
\begin{equation}
h'(z) \approx {1\over 2} z \ln z\;.
\end{equation}
 Im$Z'$ is defined by
\begin{eqnarray}
L_{int} &=& (2\sqrt 2 G_F)^{1/2}
(am_b\bar t_L b_R + b m_t\bar t _R b_L)\chi^+\;\\
\hbox{Im} Z' &=& 2\hbox{Im}(ab^*)\;.\nonumber
\end{eqnarray}
We have
\begin{equation}
\hbox{Im} Z' = 2{m_t\over m_b}{\sin2\theta\over \cos^22\theta}
\,\hbox{Im}(V_{Rtb}V_{Ltb}^*
e^{i\delta})\;, \end{equation}
and in PMLR models,
\begin{equation}
\hbox{Im} Z' = 2{m_t\over m_b} {\sin2\theta\over \cos^2 2\theta}
\sin(\delta +\alpha_b -
\alpha_t )\;.
\end{equation}
The neutron EDM from this contribution is
\begin{eqnarray}
D_n = \left \{ \begin{array}{ll}
2.5\times 10^{-27} \sin(\delta+\alpha_b-\alpha_t ) \hbox{ecm}\;,
\;\;& m_\chi = 10 \hbox{TeV}\;,\\
10^{-25}\sin(\delta +\alpha_b -\alpha_t ) \hbox{ecm}
\;,\;\; &m_\chi = 1 \hbox{TeV}\;.
\end{array} \right.
\end{eqnarray}
This result is also valid for the special case of Eq.(13).

Several comments about our results are in order:
\begin{enumerate}
\item
It is clear from our discussion that the neutral Higgs contributions to the
neutron
EDM in PMLR models are small, while the charged Higgs contributions can
be as
large as the experimental upper bound. The one loop contribution from the
charged
Higgs is smaller than the two loop contribution. However
QCD sum rule calculations show that
the the dimensional analysis estimate for the $O^C_g$ contribution may be
overestimated \cite{bu} and
the contribution from  $f_q$
may be larger than the $SU(6)$ prediction \cite{kw}.
In this case, the contribution from the charged Higgs
at the one loop level may be as important as the two loop
contribution.

If $V_L$ and $V_R$ are independent from each other, the neutral
Higgs masses can be smaller. The contribution to the neutron EDM
can then
be close to the experimental upper bound.

In MLR models, because
$V_L = V_R$ and $\delta = 0$ all the contributions discussed
above are equal to zero if there is no CP violating couplings in the Higgs
potential. We have mentioned before that in general such couplings exist. In
this case even $\delta = 0$ exchange of Higgs particle will
violate CP. The calculations are similar to those discussed before. One only
needs to change the CP violating phases in the previous equations to the CP
violating mixing parameters in this case.
The Higgs contributions to the neutron EDM are similar to those in
PMLR models.
\item   Many calculations for the neutron EDM in Left-Right
symmetric models have
concentrated on the contrbutions from $W_L-W_R$ mixing. All these contribtuions
are proportional to the mixing angle $\zeta$.  A large contribution can be
obtained from a four quark operator generated by exchange of the light $W$
boson at the tree level.  This was estimated in ref.\cite{19} to be
$$D_n \approx 2\times 10^{-19} \zeta \,{\rm Im}(V_{Lud}V_{Rud}^*)\;.$$
It is interesting to note that unless there are fortuitous cancellations,
$\zeta$ is bounded from experimental data on $\epsilon'/\epsilon$ to be less
than $10^{-5}$ if the CP violating phase involved is close to one \cite{9}.
In that case this
contribution will be smaller than   the charged Higgs contribution if
the charged Higgs mass is less than $1\;\hbox{TeV}$ and the phases of
$V_{Rtb}V_{Ltb}^*e^{i\delta}$
and $V_{Lud}V_{Rud}$ are the same order of magnitude.
\item
 Exchange of Higgs particles in Left-Right symmetric models will also generate
CP violating electron-nucleon and nucleon-nucleon interactions which will
induce
a non-zero atomic EDM. The electron-nucleon
interactions will be generated by exchange of neutral Higgs at the tree
level. We find that these interactions are small \cite{20}
($c_S$, $c_P < 10^{-10}$).
The contribution to CP violating nucleon-nucleon
interactions due to the operator $O^C_g$ from the charged Higgs are the
largest contributions due to Higgs bosons.  However it is also
very small \cite{21} ($\eta < 10^{-4}$).

\end{enumerate}

To summarise, we have studied the neutron EDM due to Higgs exchange in
Left-Right symmetric models. We find that in PMLR models
the most important effect is from the
charged Higgs at the two loop level.
In NMLR models, the neutral and charged Higgs contributions
at the one loop level can reach the experimental upper bound.
These contributions can be more important than the contributions from
$W_L-W_R$ mixing.
\acknowledgments
XGH would like to thank K.S. Babu for useful discussions.
This work was initiated at the Department of Energy's Institute for Nuclear
Theory, Seattle, which XGH, BMcK and DW thank  for its hospitality.
XGH also thanks the
University of Illinois at Chicago for hospitality where part of this work was
done. The  partial support of
the  US Department of Energy, and of the Australian Research Grants Committee,
are gratefully acknowledged.

\figure{One loop contribution to $d_{d,s}(f_{d,s})$ due to the neutral Higgs
bosons $H_{2,3}$. The $m_t^2m_b$ dependence comes from the couplings in Eq.(16)
and the mass $m_b$ insertion in the internal $b$ quark line.}
\figure{One loop contribution to $d_{d,s}(f_{d,s})$ due to the charged Higgs
boson $\chi^+$. The $m_t^3$ dependence comes from the couplings in Eq.(16)
and the mass $m_t$ insertion in the internal $t$ quark line.}
\figure{Leading two loop contribution to the quark color edm due to the
neutral Higgs exchange and the virtual top quark loop effect.}
\figure{Leading two loop contribution to the gluon color edm due to the
charged Higgs boson exchange.}

\begin{references}
\bibitem{note}For example, to get rid of the term $H_L^{\dagger} \phi H_R$
one can impose two discrete symmetries: $S_L$, which changes only $H_L$ by a
minus sign, and $S_R$, which changes only $H_R$ by a minus sign.
$S_L$ and $S_R$ form a nonabelian discrete group together with the left-right
discrete symmetry S.
\end{references}
\end{document}